\documentclass[12pt]{iopart}

\usepackage{comment}
\usepackage{cite}
\usepackage{graphicx}
\usepackage{dcolumn}
\usepackage{bm}
\usepackage{hyperref}
\usepackage[export]{adjustbox}
\usepackage[T1]{fontenc}
\usepackage{float}
\usepackage{todonotes}
\usepackage[normalem]{ulem}
\usepackage{color,xcolor}
\hypersetup{colorlinks=true,citecolor=blue,linkcolor=blue,filecolor=blue,urlcolor=blue,pdfstartview=FitH,pdfpagemode=UseNone}

\begin{document}

\title{Decoding the conductance of disordered nanostructures: a quantum inverse problem}

\author{S. Mukim$^{1,2}$, J. O'Brien$^{1,2}$, M. Abarashi$^3$, M. S. Ferreira$^{1,2}$, and C. G. Rocha$^{3,4*}$}

\address{$^1$ School of Physics, Trinity College Dublin, Dublin 2, Ireland}
\address{$^2$ Advanced Materials and Bioengineering Research (AMBER) Centre, Trinity College Dublin, Dublin 2, Ireland}
\address{$^3$ Department of Physics and Astronomy, University of Calgary, 2500 University Drive NW, Calgary, Alberta T2N 1N4, Canada}
\address{$^4$ Hotchkiss Brain Institute, University of Calgary, 3330 Hospital Drive NW, Calgary, Alberta T2N 4N1, Canada}
\ead{claudia.gomesdarocha@ucalgary.ca}
\vspace{10pt}
\begin{indented}
\item[]August 2021
\end{indented}

\begin{abstract}

Obtaining conductance spectra for a concentration of disordered impurities distributed over a nanoscale device with sensing capabilities is a well-defined problem. However, to do this inversely, i.e., extracting information about the scatters from the conductance spectrum alone, is not an easy task. In the presence of impurities, even advanced techniques of inversion can become particularly challenging. This article extends the applicability of a methodology we proposed capable of extracting composition information about a nanoscale sensing device using the conductance spectrum. The inversion tool decodes the conductance spectrum to yield the concentration and nature of the disorders responsible for conductance fluctuations in the spectra. We present the method for simple one-dimensional systems like an electron gas with randomly distributed delta functions and a linear chain of atoms. We prove the generality and robustness of the method using materials with complex electronic structures like hexagonal boron nitride, graphene nanoribbons, and carbon nanotubes. We also go on to probe distribution of disorders on the sublattice structure of the materials using the proposed inversion tool.

\end{abstract}

%
%
%
%
%

\section{Introduction}

It is a simple Quantum Mechanics problem to find the electronic conductance of a nanoscale device by directly solving the Schr\"odinger equation that governs the wave function of the quantum system under study. Obviously, this is the case if and only if the underlying Hamiltonian is known, i.e., if all scattering sources and interactions are fully specified. Trying to perform the same task in reverse is significantly more challenging. For example, assuming that the conductance of the system is known, how can one infer about the Hamiltonian components from that information alone? To make matters worse, what if the system is made of a heavily disordered material? Questions of this type are generally labeled as Inverse Problems (IP) and are classified as those which attempt to obtain from a set of observations the causal factors that generated them in the first place. IP are integral parts of classical visualization tools \cite{medical, fwi, tromp2008spectral, sonar} but far less common in the quantum realm. The literature on the field of Quantum IP (QIP) is primarily focused on the fundamentals of inversion processes, e.g., whether a problem is ill-posed \cite{Lassas2008}, whether solutions are unique and how stable they are \cite{PhysRevLett.111.090403}. Following a different scope, studies of nanomaterials and their physical properties could benefit from applications of QIP since they involve investigating structures for which the underlying Hamiltonians are not always known \cite{PhysRevX.8.031029,inv-CMP,tsymbal,jasper,gianluca,PhysRevLett.97.046401,Franceschetti1999,liping,plasma}.

Identifying the precise Hamiltonian that generates a specific observable is a difficult task. In general, it consists of solving the Schr\"odinger equation with a Hamiltonian containing one (or more) parameter(s) that must be changed until the solution closely matches the original observation. Because of such a large variance in phase space, finding the optimum parameter set can be computationally demanding. With the advance of high-performance computing and various optimization algorithms available, distinct methods for probing the parameter phase space in electronic structure problems can be used. Neural-network-based search engines \cite{jensen,yazyev}, genetic algorithms \cite{Zhang2013, Luo, Vmlinar}, and more recently, machine-learning strategies \cite{Vargas2019, kyriienko, anatole, fazli, burak, collins, Xia2018, dral, melko} have been proposed and, with different degrees of success, can speed up the search for the ``inverted'' solution. As a result, simulations are starting to have an impact in reducing the time and cost associated with materials design, specially those involving high throughput studies of material groups \cite{Ziletti2018, rajan, suram, Koinuma2004, choi, nardelli, PhysRevLett.108.068701, Yan2015,Fischer2006, Gautier2015}. These are large-scale simulations that generate volumes of data with the intention of identifying optimal combinations that can be subsequently used as candidates for an exploratory search for new materials. These methods certainly serve their purposes of determining the optimum parameter phase space that fits a given observable. Nonetheless, they can be less intuitive to implement due to their data-science scope and somehow ``black-box'' approach embedded in their algorithms \cite{Schmidt2019}. A significant portion of the electronic structure community studying low dimensional materials is certainly familiar with the Hamiltonian construction in the Schr\"odinger equation but also with its reciprocal form, involving the use of equations of motion for the inverse of the Hamiltonian operator which can be identified as the Green's function. Green's function-based methods\cite{abcd} have played a central role as a theoretical tool to study electronic and transport properties of nanoscale systems. These are versatile methods that can be used to obtain the conductance of complex low dimensional materials, including disordered ones.

A mathematically transparent inversion technique capable of extracting structural and composition information from a disordered quantum device by looking at its conductance signatures has recently been proposed \cite{shardul}. Using energy-dependent two-terminal conductance as input functions, the reported inversion method identifies the concentration of impurities disorderly distributed within a graphene nanoribbon (GNR) host. Despite claims that the method is general, robust, and stable, evidence has so far been limited to nitrogen-doped GNRs \cite{shardul}. In this manuscript, we put the generality of this inversion methodology to the test by implementing it with a variety of systems, some described by extremely simple electronic structure models, others by more realistic ones. Furthermore, the inversion accuracy is also tested and shown to be able to identify correctly the impurity concentration of a number of systems studied in this work. This suggests that this method can indeed be employed as a useful characterisation tool with a variety of systems composed of different materials and dimensions. The method relies on working with an objective function that can be built directly from conductance calculations which we call the `misfit' function. A remarkable advantage of the method is its systematic approach in dealing with the misfit function, without the need of extra adjustments or parameter tuning to perform the optimization search of concentration and impurity characteristics. This article starts by illustrating in the next section the inversion methodology applied to a very simple one-dimensional atomic chain system. In subsequent sections, we demonstrate how this inversion tool can be implemented to extract valuable information about nanoscale systems acting as hosts of chemical impurities; hosts include carbon nanotubes (CNTs), GNRs, and hexagonal boron nitride (hBN) nanoribbons, on which foreign atoms are randomly distributed over their atomic structure. 

\section{Methodology}

To demonstrate the versatility of this inversion methodology, we separate the cases we study into two groups: one with systems possessing very simple electronic structures and another in which the materials are described by more realistic band structures. The purpose of separating the case studies is twofold: (i) it is a lot easier to introduce the inversion procedure with simple systems since it helps preventing artifacts coming from complex electronic structure models and our focus lies entirely on the inversion method itself; (ii) showing that the inversion method works for systems regardless of their electronic-structure details and of their dimensionality is strong evidence of the robustness of the methodology. With that in mind, we will present the inversion method with the first group consisting of three different one-dimensional systems, all possessing rather simple electronic structures. The first system consists of an electron gas in the presence of short-range scatters represented by randomly distributed delta functions. In this case, the delta-functions act as impurities in the presence of an otherwise homogeneous potential produced by an infinitely-long host material. Mathematically, the single-particle Hamiltonian is defined by the following potential
\begin{equation}
    V_1(x) = \sum_{j=1}^N \lambda \, \delta(x-x_j) \,\,,
    \label{delta-potential}
\end{equation}
where $j$ is an integer running from $1$ up to the total number of scatters $N$, $x$ is the spatial variable, $x_j$ gives the randomly-selected position of the $j^{th}$ scatter, and $\lambda$ is a positive constant that reflects the scattering strength of the impurities, assumed to be all identical for simplicity. The $N$ random values of $\{x_j\}$ are constrained to be within a region of length $L$ and fully determine the Hamiltonian as well as the corresponding transmission coefficient $T(E)$, $E$ being the energy. Note that the integer number of impurities $N$ or the non-integer concentration $n=N/L$ are considered equivalent control parameters and will be used interchangeably throughout this article. 

In the second case study considered here, we relax the short-range aspect of our scatters by replacing the delta-function potentials with identical square barriers of width $d$ and height $V_0$. In other words, the new Hamiltonian has now a potential described by  
\begin{equation}
    V_2(x) = \sum_{j=1}^N V_0 \, [\Theta(x-x_j) - \Theta(x-x_j-d)] \,\,,
    \label{barrier-potential}
\end{equation}
where $\Theta$ is the Heaviside step function. Once again, the barrier positions are determined by randomly selected values of $\{x_j\}$ with the additional constraint that barriers must not overlap, {\it i.e.}, no pair of values $x_j$ and $x_{j^\prime}$ must be less than a distance $d$ apart.

The third and final case studied departs from a free-electron gas description for the electronic structure of the host material. Still keeping a simplified picture of the electronic structure, we move on to a lattice representation of the Hamiltonian, in one-dimension, captured by the tight-binding picture of an infinitely long chain of atoms. With no loss of generality, we adopt the following Hamiltonian  
\begin{equation}
    {\hat H}_3 =  \sum_j  ( \, \vert j \rangle t \langle j+1 \vert + \vert j \rangle t \langle j-1 \vert \, ) + \sum_{j^\prime} \vert j^\prime \rangle \lambda_{j^\prime} \langle j^\prime \vert \,\,,
    \label{tb}
\end{equation}
which contains an off-diagonal hopping term and a diagonal term describing the potential on all sites of the system. In equation (\ref{tb}), $j$ and $j^\prime$ are integers that label the atomic sites, $\vert j \rangle$ represents an electronic orbital centred at site $j$ and $t$ is the electronic hopping between nearest-neighbour sites only. $\lambda_{j^\prime}$ represents the on-site potential of substitutional impurities that are once again randomly distributed within a region of size $L$. It is worth mentioning that impurities in the form of adatoms (adsorption) can also be considered in the Hamiltonian in equation (\ref{tb}) without fundamental changes to our approach. 

\begin{figure}
    \centering
    \includegraphics[height=0.57\columnwidth,right]{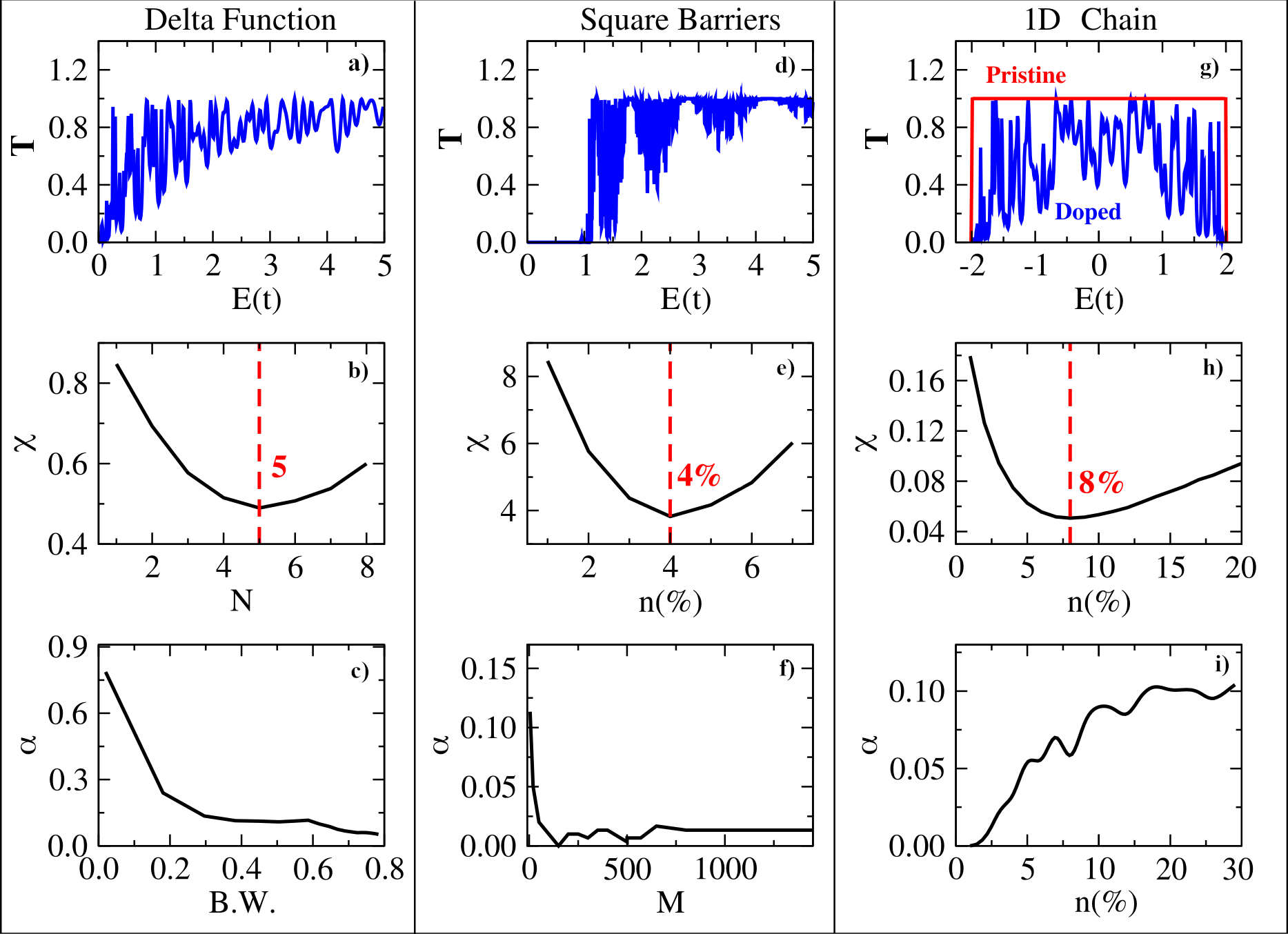}
    \caption{(first row) Transmission coefficient ($T$), (second row) misfit function ($\chi$), and (third row) error ($\alpha$) as a function of representative physical quantities in disordered systems: $E$ is the energy expressed in certain units of energy $t$, $N$ is the number of scatters, $n$ is the impurity concentration expressed in $\%$, B.W. is the normalized bandwidth used for integration in the inversion procedure, and $M$ is the total number of disordered configurations in the studied ensemble. These results were obtained for three case study systems: (first column) electron gas with delta functions representing scatters, (second column) electron gas with identical square barriers of width $d=1$ and height $V_0=1$ in arbitrary units, (third column) one-dimensional (infinite) atomic chain. In all cases, $t=1$. For the third case (atomic chain), the transmission is scaled in terms of the conductance quantum, $\Gamma_0 = 2e^2/h$ with $e$ being the elementary charge, and $h$ the Planck constant. Panel g) also contains the transmission as a function of energy for the pristine atomic chain (red curve). Impurities in the doped case were modelled with on-site energy of $\lambda = 0.5t$ and they were spread within a length of $L=100$ atomic layers. Transmission results in the first row are for the target systems in which the misfit function locates a minimum: (a,b) electron gas with $N_{min}=5$ delta functions, (d,e) electron gas with $n_{min}=4\%$ square barrier scatters, and (g,h) 1D atomic chain with $n_{min}=8\%$ impurities. }
    \label{fig-simple}
\end{figure}

All three cases introduced above can be characterised by an infinitely-long host containing a number $N$ of randomly distributed impurities confined to within a finite-sized region and described by the respective equations (\ref{delta-potential}), (\ref{barrier-potential}), and (\ref{tb}). With this general setup, one can obtain the transmission coefficient across all three systems following a method of choice, e.g. Green function\cite{abcd,rocha1,rocha2,lawlor1} or scattering matrix\cite{datta_1995}. The top three panels of Fig.~\ref{fig-simple} depict the electronic transmission $T(E)$ for a specific configuration of each one of these simple cases. It is worth noting that in each one of the three cases, the exact number of impurities and their respective locations defining the parent Hamiltonian that generated the $T(E)$ curve of Fig.~\ref{fig-simple} are set aside and will not be used in any part of the subsequent inversion calculation. In other words, $T(E)$ serves as the only input function from which the inversion will take place. The task at hand is a QIP that aims to find structural and composition information about the scattered impurities using the seemingly noisy curves shown on the top panels of Fig.~\ref{fig-simple} as the only starting point. The transmission coefficient is a rather representative quantity to use as input function because it has several parallels in real materials that go beyond this simple illustrative model. For example, the conductance of a metal is commonly expressed in terms of $T(E)$, following the Landauer-B\"uttiker formalism \cite{datta_1995}. Likewise, other quantities such as thermal and optical conductivity can also be expressed in a spectral representation that resembles the transmission coefficient\cite{tuovinen}. In this work, we put emphasis on extracting materials' composition from electronic transport analysis following the misfit function minimization scheme as illustrated in the case studies above. It is important to emphasize a general aspect of our methodology which is the possibility of decoding not only electronic conductance curves but also response functions of other physical natures, e.g. mechanical, optical, magnetic, etc. For instance, this methodology was recently adapted to decode the optical conductivity of disordered MoS$_2$ films as presented in Duarte et al. \cite{duarte2021}. Another promising applicability of the methodology is to use in the context of mechanical response in line with the approach proposed by Bao et al. \cite{bao2013}. Note also that our misfit function methodology is not limited to conductance curves derived from numerical methods; our method is applicable to any energy-dependent response function that can serve as a ``target'' function for the misfit function minimization scheme. Typical experimental data that could be used with this methodology could come from Hall-bar experiments, in our case, conducted in 2D nanostructures. As long as the main methodology parameters (i.e., bandwidth, size of the ensemble, transport regime, and hamiltonian flavor) are set to achieve the desired accuracy, a convergence to a minimum in the misfit function can be expected for any fluctuating energy-dependent response function obtained from experiments or numerical means.  

As previously alluded to, a naive approach would be to attempt a comparison of the input function with the transmission coefficient of every single possible disorder configuration but the prohibitively large number of combinations proves this strategy impracticable.

Fortunately, the inversion procedure introduced in \cite{shardul} provides a far more
efficient technique which combines configurationally-averaged (CA) values of the transmission coefficient with the ergodic principle. More specifically, the ergodic hypothesis assumes that a running average over a continuous parameter upon which the transmission coefficient depends is equivalent to sampling different impurity configurations and this can be used to assist us in the inversion procedure. In mathematical terms, this is done through the so-called misfit function defined as \cite{shardul}
\begin{equation}
\chi(n)={1 \over {\cal E}_+ - {\cal E}_-}\,\int_{{\cal E}_-}^{{\cal E}_+} dE \,  (T(E) - \langle T(E, n) \rangle)^2\,\,,
\label{misfit}
\end{equation}
where the integration limits are arbitrary energy values. $\chi$ can be interpreted as a functional that measures the deviation between the input transmission of the parent configuration $T(E)$ and its CA counterpart  $\langle T(E) \rangle$, defined as
\begin{equation}
\langle T(E,n) \rangle = {1 \over M} \sum_{m=1}^M T_m(E).
\label{CA}
\end{equation}
where $M$ is the total number of different disordered configurations taken into account and $T_m(E)$ is the corresponding transmission coefficient for each one of these configurations, which are themselves labelled by the integer $m$. While both $T(E)$ and $\langle T(E) \rangle$ in the integrand above are functions of energy, the latter is also a function of the impurity concentration $n$. When plotted as a function of $n$, the misfit function displays a very distinctive minimum at a value that corresponds to the real impurity concentration. This can be seen in the middle panels of Fig.~\ref{fig-simple}. The concentration $n_{min}$ that minimises the misfit function coincides with the real concentration $n_i$ in all three cases, indicating that the inversion is successful. Also included in the Figure (bottom-row panels), we plot the relative error $\alpha = \vert n_{min} - n_i\vert/n_i$, where $n_{min}$ stands for the concentration for which the misfit function is minimum. The error $\alpha$ is calculated over several different parent configurations in order to achieve statistical significance.

One question that naturally arises is how relevant the energy integral of equation (\ref{misfit}) is. On the one hand, the energy integration plays a significant part in identifying the real impurity concentration because $T(E)$ fluctuations of a single sample versus energy is equivalent to sample-to-sample fluctuations at a fixed energy. Therefore, this energy integration is similar to vastly augmenting the number of disordered configurations taken into account. In fact, if we were to use equation (\ref{misfit}) with ${\cal E}_-={\cal E}_+$, the misfit function would not have any distinctive feature and would not lead to a successful inversion \cite{shardul}. However, if the integration limits span a small fraction of the relevant energy range, the misfit function acquires a very distinctive shape with minima located at the correct concentrations. 

Some further insights we can learn from Fig.~\ref{fig-simple} are as follows. Fig.~\ref{fig-simple}(c) plots the inversion error $\alpha$ as a function of the normalized bandwidth defined as B.W.$=({\cal E}_+ - {\cal E}_-)/Z$, with $Z$ being the total bandwidth of the electron gas system. Note that $\alpha$ is large for very narrow energy windows. When the energy window for integration is broadened beyond $20\%$ of the bandwidth, $\alpha< 0.1$ in the case of $N=5$ delta function potentials. This indicates that larger energy windows accounted in the misfit function can capture more of the transmission features, minimizing then the error of the inversion procedure. The integration in equation (\ref{misfit}) can
be hence replaced with a discrete sum without major impact to the accuracy of the
inversion procedure. By reducing the number of energy points ${\cal N}_\epsilon$ involved in generating the misfit function, there is a point below which the inversion accuracy drops quite significantly but this threshold is normally at very low levels (${\cal N}_\epsilon < 50$), which enables calculations with systems for which the numerical complexity is significantly higher \cite{shardul}, as we will demonstrate in the next section. 

Regarding the error study done in Fig.~\ref{fig-simple}(f) and the number of configurations needed in the CA values $M$, the backbone of our inversion procedure is that fluctuations in the conductance ($\Gamma$) contain very little system specific information but the average conductance depends smoothly on the variable of interest. In this sense, the higher the value of disorder realizations $M$, the smaller the fluctuations in $\langle \Gamma(E,n) \rangle$. Universal conductance fluctuations (UCF) can be used to estimate values of $M$ required for sufficiently accurate results in the inversion procedure. Both for diffusive \cite{lee1985universal,lee1987universal} and chaotic ballistic systems \cite{Baranger1994}, var($\Gamma) \approx 1$ is the main fingerprint of UCF. Hence, the CA relative statistical error is expected to scale with $[\sqrt{M}\times \langle \Gamma(E,n) \rangle]^{-1}$. To bring the degree of fluctuations to an acceptable level that yields statistically significant results, we estimate that $M$ should be of the order of $10^3$ \cite{Alhassid2000,lopez2014modeling}. As seen in Fig.~\ref{fig-simple}(f), the error is only guaranteed to stabilize at $M\sim 1000$. Depending on the complexity of the system, higher values of $M$ can be required to achieve statistical significance.

Finally, Fig.~\ref{fig-simple}(i) shows the inversion error $\alpha$ for an infinite 1D atomic chain as a function of impurity concentration $n$. Substitutional impurities were spread randomly over $L=100$ unit cells in length. One can see that $\alpha$ is relatively small for concentrations below 10\%, clearly indicating that our inversion method is very reliable for dilute regimes. To test the inversion procedure, we defined hamiltonians with the knowledge of concentration as well as positions. By increasing the concentration of scatters spread over a finite length, one can start forming pairs or atomic clusters of impurities, in which the method is not equipped to distinguish from single scatters. For this reason, we see a reduction in the accuracy of our method ($\alpha$ increases) as impurity concentration increases.

\section{Results and discussion}

\begin{figure}
    \centering
    \includegraphics[height=0.57\columnwidth,right]{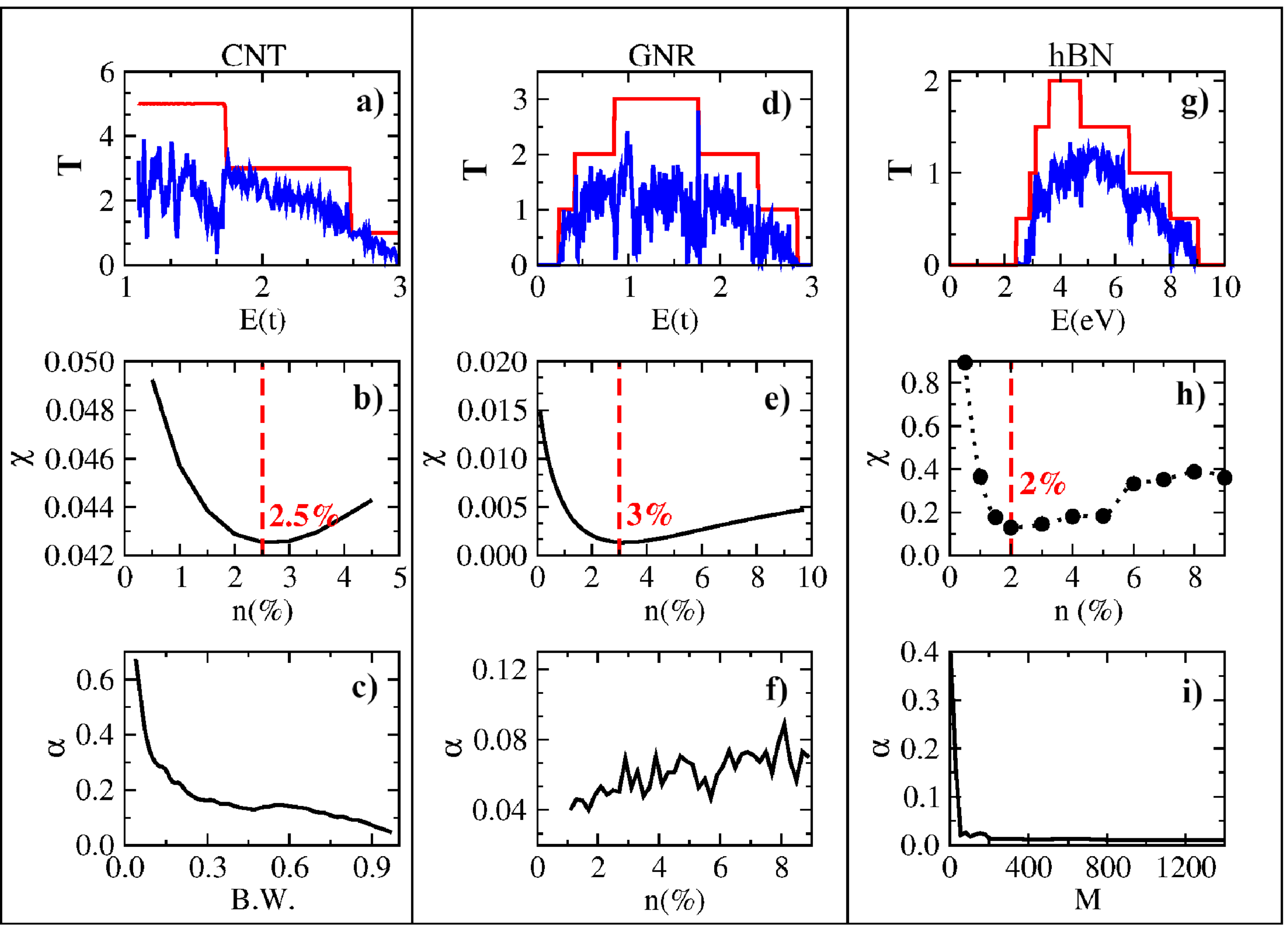}
    \caption{
    (first row) Electronic transmission ($T$), (second row) misfit function ($\chi$), and (third row) error ($\alpha$) as a function of representative physical quantities in disordered systems: $E$ is the energy expressed in units of hopping $t$ or directly in eV, $n$ is the impurity concentration expressed in $\%$, $M$ is the total number of disordered configurations in the studied ensemble, and B.W. is the normalized bandwidth used for integration in the inversion procedure. These results were obtained for three quasi-1D systems: (first column) carbon nanotube (CNT), (second column) graphene nanoribbon (GNR), and (third column) hexagonal boron nitride (hBN) nanoribbon. The transmission results on the top panels are scaled in terms of the conductance quantum, $\Gamma_0 = 2e^2/h$ with $e$ being the elementary charge, and $h$ the Planck constant. Top panels (a,d,g) contain the transmission as a function of energy for the respective pristine systems (red curves) and for the doped target systems (blue curves) in which the misfit function locates a minimum: (a,b) CNT at $n_{min}=2.5\%$, (d,e) GNR at $n_{min}=3\%$, and (g,h) hBN at $n_{min}=2\%$ impurities. The CNT is an armchair (5,5) hosting impurities modelled with an on-site energy of $\lambda = 0.5\, t$ with $t=1$ spread over L=50 unit cells in length. The GNR is an armchair-edge nanoribbon with 7 atoms along its width and it is hosting impurities with an on-site energy of $\lambda = 0.5\, t$ with $t=1$ spread over L=100 unit cells in length. The hBN host is an armchair-edge ribbon with 8 atoms along its width and infinite in length. Carbon atoms act as impurities and are spread over $L=100$ layers long. The on-site energy values to model B, N, and C atoms are: $\lambda_B = -6.64$ eV, $\lambda_N = -11.47$ eV, and $\lambda_C = -8.97$ eV\cite{dibenthesis}. See main text for details on the hopping characteristics used to model the hBN.}
    \label{fig:gnr}
\end{figure}
We now proceed to introduce three new case studies in the second group of systems, which are less simplistic and more realistic. The first system in this group consists of carbon nanotubes (CNTs); the second is made of graphene nanoribbons (GNRs); the third and final case is made of hexagonal boron nitride (hBN) nanoribbons. Each one of these materials will contain a finite concentration of substitutional impurities that differ from their pristine atomic composition. One obvious difference to the previous cases is that these are no-longer one-dimensional systems and their electronic structures are a lot more involved than the ones displayed by the cases shown in Fig.~\ref{fig-simple}.
Even though these new cases possess band structures that have finer details than the ones in the first group, it is still convenient to maintain the tight-binding description of the electronic structure. Successful inversions have already been reported with Density Functional Theory (DFT) calculations used to describe the multi-orbital electronic structure of doped graphene nanoribbons \cite{shardul}. Despite its success, {\it ab-initio} calculations are far more time-consuming than semi-empirical approaches such as tight-binding, thus justifying our choice of electronic structure model for this manuscript. Therefore, the Hamiltonian appearing in equation (\ref{tb}) is still applicable to describe the cases in this group, the main difference being that the integer $j$ now accounts for all sites of these quasi-2D structures. Particularly for the hBN case, the Hamiltonian will carry three on-site contributions to account for the nitrogen ($\lambda_N$), boron ($\lambda_B$), and impurity atoms as we will detail later on.  

Results found for these new cases are shown in Fig.~\ref{fig:gnr} with CNT on the left panels, GNR on the middle panels and hBN on the right panels. Following the same pattern of Fig.~\ref{fig-simple}, the top-row panels display the electronic transmission of the parent configuration that serves as the input function in the inversion procedure. Note that in this case, these functions can be interpreted as the Fermi-energy-dependent conductance. Also on the same plots are the impurity-free conductance plots shown for comparison purposes. The middle-row panels depict the corresponding misfit function $\chi(n)$, always plotted as a function of the concentration of impurities $n$. Once again for the sake of comparison, the same plot contains a vertical dashed line that indicates the real impurity concentration contained in the parent configuration. Finally, the bottom-row panels indicate the accuracy of the inversion procedure by plotting the error $\alpha$ as a function of certain control parameters that will be discussed later. 

For the CNT results on the left-column panels of Fig.~\ref{fig:gnr}, they contain impurities represented by on-site potentials of $\lambda= 0.5$ in units of hopping (see equation (\ref{tb})). While this value corresponds to an arbitrary choice that does not represent any specific atomic species, the inversion method correctly identifies the impurity concentration for whichever choice of $\lambda$ values. Such a good agreement between the inverted result and the real concentration is seen in the misfit function plot for the nanotube. The concentration that minimizes $\chi$, $n_{min}$, coincides with the $n_i=2.5\%$ chosen for the parent configuration. Having performed the inversion procedure with numerous different parent configurations, we were able to assess its success rate in the case of nanotubes serving as hosts. Fig.~\ref{fig:gnr}(c) depicts the error $\alpha$ plotted as a function of the bandwidth, revealing that the accuracy of the method improved as the integration window of the misfit function increased. The central-column panels display equally good results for the GNR with the same choice of $\lambda = 0.5$ in units of hopping to mimic substitutional impurity atoms. Fig.~\ref{fig:gnr}(d) shows the conductance of a pristine armchair-edged GNR with 7 atoms wide plotted as a function of energy. Note that only the conduction band ($E>0$) is shown to simplify visualization. On the same panel, the conductance of the parent configuration containing 3\% of impurities is also shown. Fig.~\ref{fig:gnr}(e) displays the $n$-dependent misfit function with a distinctive minimum that agrees with the real concentration, $n_i=3\%$. The accuracy for the nanoribbon inversions is tested by plotting $\alpha$ as a function of concentration in Fig.~\ref{fig:gnr}(f). Once more, we observe that the inversion method performs better in dilute regimes of doping, with the error $\alpha$ increasing relatively slow with the concentration of impurities. 

The right-column panels of Fig.~\ref{fig:gnr} show results for an armchair-edge hBN nanoribbon with 8 atoms along its width. The ribbon is infinite in length but, in its doped form, impurities are spread over a section of $L=100$ layers long. The tight-binding Hamiltonian for the hBN host uses three distinct on-site energy values to model B and N atoms on the host and C atoms as impurities of the system: $\lambda_B = -6.64$ eV, $\lambda_N = -11.47$ eV, and $\lambda_C = -8.97$ eV\cite{dibenthesis}. Hopping terms were parameterized as $t =-6.17/a^2$ \cite{dibenthesis, harrison} with $a$ being the pair-wise bond length. For a boron-nitrogen bond, $a=1.43$ \AA\, and, for the sake of simplicity, we consider that boron-carbon and nitrogen-carbon bond lengths do not change significantly from that value. The noisy conductance curve in Fig.~\ref{fig:gnr}(g) corresponds to the parent system of a hBN ribbon doped with $n_i=2\%$ of substitutional impurities and the red curve is the conductance of the pristine hBN nanoribbon. The misfit function shown in Fig.~\ref{fig:gnr}(h) acquires a minimum at $n_{min}=2\%$, evidencing that the method predicted correctly the concentration of the parent system. Finally, Fig.~\ref{fig:gnr}(i) confirms that the accuracy of the method is significantly improved as the number of disordered configurations in the ensemble, $M$, increases.
\begin{figure}[!tbp]
    \centering
    \begin{minipage}[b]{0.45\textwidth}
    \includegraphics[width=\textwidth,right]{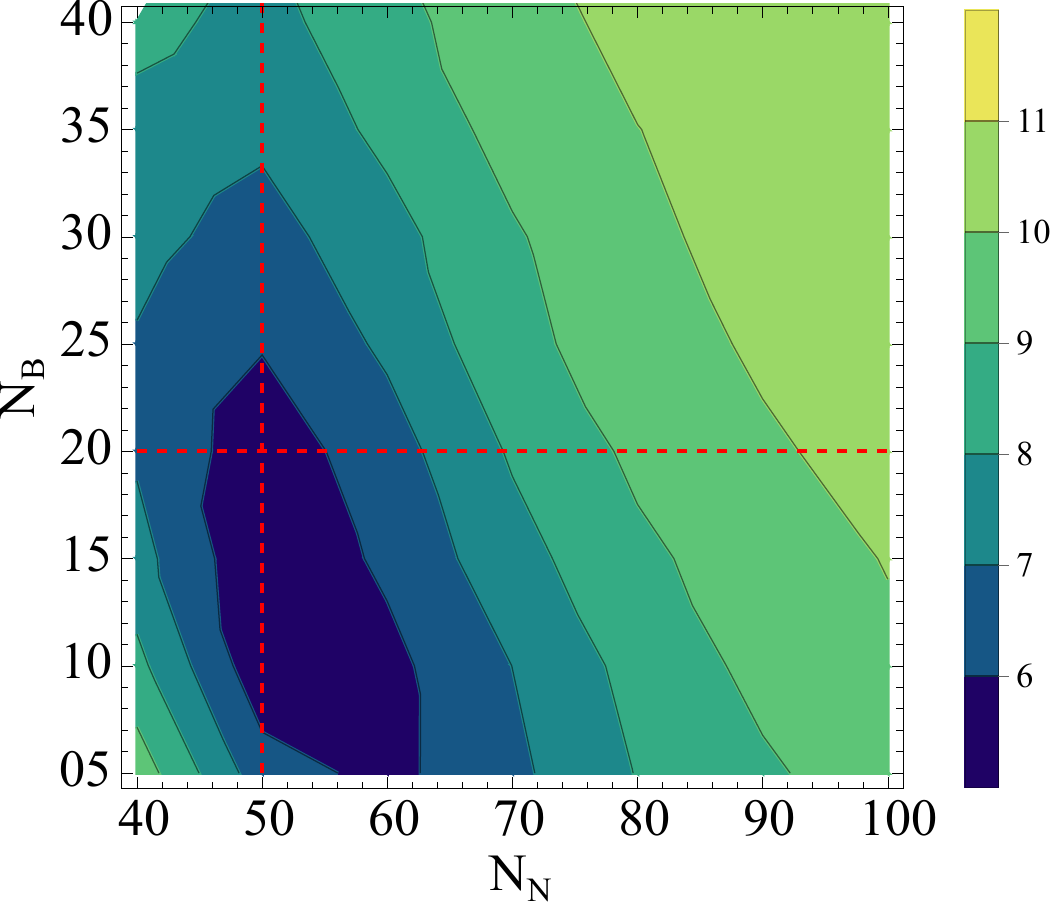}
    \end{minipage}
    \begin{minipage}[b]{0.45\textwidth}
    \includegraphics[width=\textwidth,left]{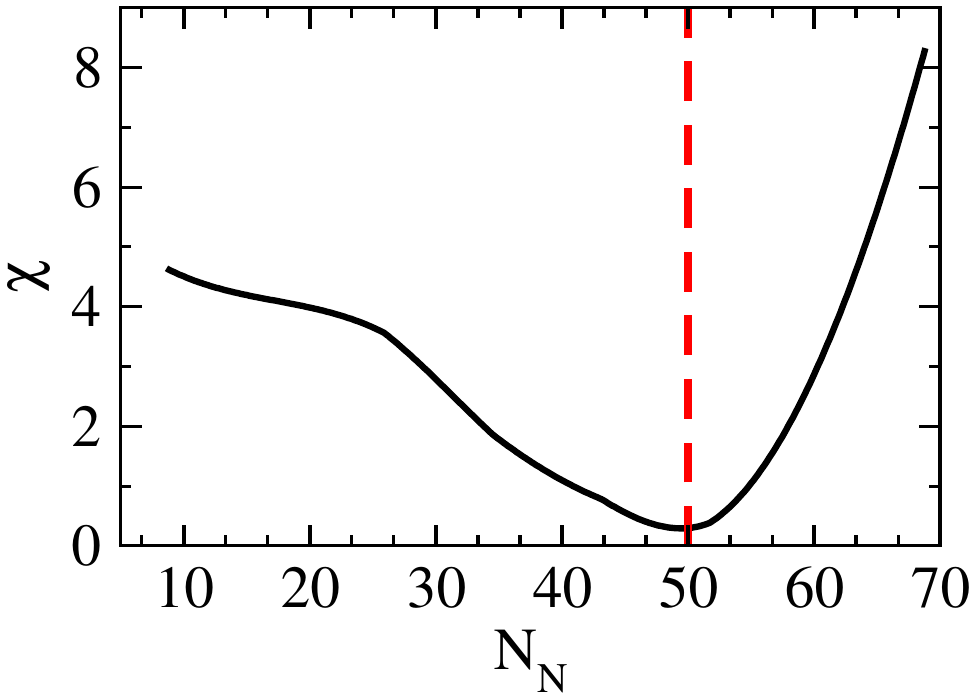}
    \end{minipage}
    \caption{(Left panel) Misfit function ($\chi$) surface plot taken for an armchair-edge hBN nanoribbon with 7 atoms along its width. The ribbon is infinite but carbon impurities were spread randomly over $L=100$ unit cells along its length. The bar color express values of $\chi$. The control parameters along x- and y-axis are $N_N$ and $N_B$, respectively; those account for the number of carbon impurities replacing nitrogen ($N_N$) or boron ($N_B$) atoms. Dashed lines intersect at the characteristics of the parent system with $N_{N} = 50$ and $N_B=20$. The minimum of $\chi$ correctly predicts the characteristics of the parent system as it lays within the region where $N_{N} = 50$ and $N_B=20$. (Right panel) Curve of $\chi$ versus $N_{N}$ taken from left panel by slicing the misfit surface plot horizontally at $N_{B} = 20$ (red horizontal dashed line). A distinctive minimum at $N_{N}=50$ (red vertical dashed line) correctly finds the occupation of carbon impurities on nitrogen sites of the parent system. Bandwidth considered for inversion is 70\% of the entire spectrum. }
    \label{Hbn_N}
\end{figure}

Another way to prove the generality of our methodology is to extend its use to a multi-dimensional phase space. Till this point, the inversion procedure looked at one degree of freedom which was the impurity concentration. However, it is possible to extend this analysis to a two-dimensional phase space, e.g., as we are going to illustrate. This analysis turns particularly interesting if applied to the hBN case because impurities can replace two types of atoms on the host (N or B). This means we can decompose the concentration information into the two distinct sublattices. Considering that a total of $N$ impurities can be substitutionally dope a segment length of an hBN nanoribbon, we can write that $N=N_B+N_N$ where $N_B$ ($N_N$) is the number of boron (nitrogen) atoms replaced by an impurity. Therefore, we can write the misfit function in terms of these two degrees of freedom to probe occupation of impurities on both boron and nitrogen sites as 
\begin{equation}
    \chi(N_{B},N_{N}) =  \int_{{\cal E}_-}^{{\cal E}_+} dE \, \left[\, T(E) - \langle T(E,N_{B},N_{N})\rangle \, \right]^2 \,\,
    \label{chi2d}
\end{equation}
A certain number of C atoms are then spread over the hBN host in such a way that they can replace equal portions of B and N atoms or they can cause a sublattice unbalance. In other words, we can define a parameter $\delta$ as $\delta = N_N - N_B$ that can serve as a metric for this unbalance: if $\delta>0$, C atoms majorly occupy the nitrogen sublattice, if $\delta<0$, C atoms majorly occupy the boron sublattice, and if $\delta=0$, C atoms occupy both sublattices equally.  
A 2D contour plot of $\chi$ as defined in equation (\ref{chi2d}) is presented in Fig.~\ref{Hbn_N} left panel. The global minimum correctly coincides with the occupation of impurities placed on boron ($N_B=20$) and nitrogen ($N_N=50$) sublattice of the parent system highlighted by the intersection of vertical and horizontal dashed red lines. The right panel in Fig.~\ref{Hbn_N} corresponds to an horizontal slice taken from the surface plot at fixed $N_B=20$ to visualize the dependency of $\chi$ with $N_N$ and to highlight the minimum of $\chi$ at $N_N=50$. The fundamental difference between boron and nitrogen sites allows to identify occupation of carbon impurities on corresponding sublattices. However, identifying the exact position of impurities still remains an elusive task. This example proves that extending the inversion methodology to inspect more than one degree of freedom does not affect its generality and robustness. Looking at more than one degree of freedom in the misfit function certainly increases computational cost and may generate multiple minima. Nonetheless, the method can serve as a first approach to determine the exact features of a parent system that can be described in a multi-dimensional phase space. In other words, results as the one depicted in Fig.~\ref{Hbn_N} narrows down considerably the range of search for the optimum $N_B$ and $N_N$ values. A refinement in the range of values can be achieved by re-applying the inversion procedure within the reduced parameter range and with an increased number of CA samples, $M$, and/or by increasing the bandwidth established for the inversion procedure.

It is worth mentioning that our methodology is not immune to the typical limitations regarding the investigation of large-scale systems containing over $10^3$ atoms. Yet, we tested our methodology with relatively large systems, e.g. nanoribbon systems with $\sim 2000$ atoms along their length. A single point calculation to determine the accuracy of the method took approximately 30-60 minutes to complete depending on the ensemble size. Depending on the computational resources available, this computing time can be significantly reduced. Certainly systems treated with more complex Hamiltonians, beyond the nearest-neighbour single-orbital tight-binding approximation, can take longer computational times. Nonetheless, our methodology can be paired with other scalable electronic transport approaches, e.g. as the one proposed by Fan et al. \cite{FAN201428} in which an optimization for the linear-scaling Kubo-Greenwood formula is presented, allowing large-scale calculations in systems of $\sim 10^6$ atoms.

\section{Conclusions}
In this manuscript, we illustrated the use of an inversion methodology capable of extracting information out of disordered systems, more specifically host media perturbed by local potentials representing impurities or dopants. For situations in which we know the initial conditions of the system of study, the method may sound unimpressive. However, if we do not have access to its initial conditions, one needs to find a way of retrieving the system initial setup by inspecting response functions of the system subjected to perturbations. In this work, we studied a series of solid state systems ranging from electron gas, 1D atomic chain, up to carbon nanotubes, graphene and hBN nanoribbons that give the conductance or electronic transmission as response function. Each of these (host) solid state systems were perturbed by a certain concentration of impurities and their presence induce fluctuations in the transmission response functions, turning them into quite noisy spectra. It is not straightforward to deduce the concentration of impurities doping the hosts by only looking at the noisy profile of the transmission curve. We proved the generality of an inversion methodology proposed by Mukim et al. \cite{shardul} in which unknown quantities such as the impurity concentration in doped nanoscale systems can be uncovered by computing the so called ``misfit function'', an objective function written in terms of the configurationally averaged conductance and the target conductance. The misfit function reveals minima at locations in the phase space parameter that correspond to the unknown initial conditions of the system, in this case, uniquely characterized by the concentration of impurities doping a host material. The method itself can be built upon numerous control parameters that inform on its accuracy and performance such as relative error to target, influence of bandwidth window, number of configurationally averaged samples in the ensembles, to name but a few. We observed that the accuracy of the method improves considerably when: (i) sufficiently large bandwidths are selected for the integration of the misfit function, (ii) the sampling of the ensemble is increased, (iii) target systems are in dilute regime of sufficiently low concentration of dopants. In summary, the method proved capable of ``decoding'' noisy transmission response functions into a more meaningful mathematical representation, the misfit function, that exhibits minima at the characteristics defining the unknown a priori initial conditions of the studied systems.  

\section*{Acknowledgements}
This publication has emanated from research supported in part by a research grant from Science Foundation Ireland (SFI) under Grant Number SFI/12/RC/2278-P2. This work was also supported by U of C start-up funding and partially support by the Natural Sciences and  Engineering Research Council of Canada (NSERC), [Discovery Grant funding reference number xxxxxx]. We also acknowledge the WestGrid (www.westgrid.ca), the Compute Canada Calcul Canada (www.computecanada.ca), and the CMC Microsystems (www.cmc.ca) for computational resources. 

\

\

\bibliography{mybib,dftbib,newref}

\end{document}